\documentclass[aps,pra,reprint,showpacs,floatfix,longbibliography]{revtex4-1}
\usepackage{graphicx, amsmath, amssymb}
\usepackage[caption=false]{subfig}
\allowdisplaybreaks[1] 
\begin{document}


\newcommand{\braket}[2]{{\left\langle #1 \middle| #2 \right\rangle}}
\newcommand{\bra}[1]{{\left\langle #1 \right|}}
\newcommand{\ket}[1]{{\left| #1 \right\rangle}}
\newcommand{\ketbra}[2]{{\left| #1 \middle\rangle \middle \langle #2 \right|}}


\title{Quantum Search with Multiple Walk Steps per Oracle Query}

\author{Thomas G.~Wong}
	\email{twong@lu.lv}
\author{Andris Ambainis}
	\email{andris.ambainis@lu.lv}
	\affiliation{Faculty of Computing, University of Latvia, Rai\c{n}a bulv.~19, R\=\i ga, LV-1586, Latvia}

\begin{abstract}
	We identify a key difference between quantum search by discrete- and continuous-time quantum walks: a discrete-time walk typically performs one walk step per oracle query, whereas a continuous-time walk can effectively perform multiple walk steps per query while only counting query time. As a result, we show that continuous-time quantum walks can outperform their discrete-time counterparts, even though both achieve quadratic speedups over their corresponding classical random walks. To provide greater equity, we allow the discrete-time quantum walk to also take multiple walk steps per oracle query while only counting queries. Then it matches the continuous-time algorithm's runtime, but such that it is a cubic speedup over its corresponding classical random walk. This yields the first example of a greater-than-quadratic speedup for quantum search over its corresponding classical random walk. 
\end{abstract}


\maketitle


\section{Introduction}

Quantum walks are the quantum analogues of classical random walks \cite{Ambainis2003,Kempe2003}, and they have been the subject of much investigation for their algorithmic role in search \cite{SKW2003}, element distinctness \cite{Ambainis2004}, triangle finding \cite{MSS2005}, and evaluating NAND trees \cite{FGG2008}. Formulated on a graph, a quantum particle walks locally from one vertex of the graph to adjacent vertices in superposition.

As with classical Markov chains, quantum walks can evolve in discrete or continuous time. But these two formulations differ in the required number of degrees of freedom. With the $N$ vertices of a graph labeling computational basis states of an $N$-dimensional Hilbert space, continuous-time quantum walks are well-defined on these vertices. Discrete-time quantum walks, however, require additional ``coin'' or spin degrees of freedom in order to evolve non-trivially \cite{Meyer1996a,Meyer1996b}. This leads to some algorithmic differences between the two approaches \cite{CG2004,AKR2005}, and much work has been done to show the relationship between them \cite{Meyer1996a,CG2004b,Strauch2006,Childs2010,DB2015}.

In this paper, we identify a key difference between discrete- and continuous-time quantum walks as they are typically used in solving spatial search \cite{AA2005}, where the goal is to find a ``marked'' vertex in a graph by querying an oracle. The oracle is considered to be an expensive ``black box'' that we want to use as little as possible, and other operations are ``cheap.'' Following this standard for oracular problems, we compare algorithms by their oracle query complexity. As we will show in the following sections, the usual discrete-time quantum walk search algorithm takes one walk step per oracle query, while the standard definition of the continuous-time algorithm effectively allows multiple walk steps per oracle query.

\begin{figure}
\begin{center}
	\includegraphics{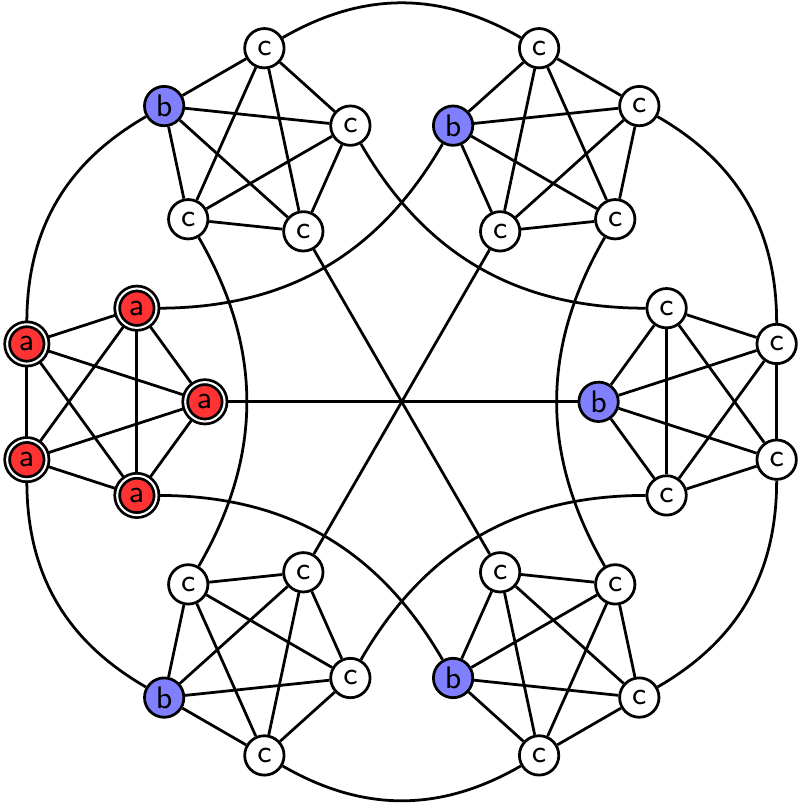}
	\caption{\label{fig:simplex_full} A 5-simplex with each vertex replaced by the complete graph of 5 vertices. One complete graph is fully marked, indicated by double circles. Identically evolving vertices are identically colored and labeled.}
\end{center}
\end{figure}

We show how this difference affects search on the ``simplex of complete graphs,'' which was first introduced in \cite{MeyerWong2014}, and an example of which is shown in Fig.~\ref{fig:simplex_full}. In the graph, we have $M+1$ complete graphs of $M$ vertices, arranged so that so that the vertices of a complete graph are each connected to different complete graphs. Here, we search for one fully marked complete graph, as indicated by the red $a$ vertices with double circles in Fig.~\ref{fig:simplex_full}. We also have the blue $b$ vertices that are one away from (\textit{i.e.}, adjacent to) the $a$ vertices, and the white $c$ vertices that are two away from the $a$ vertices. Classically, a random walker is expected to transition from one complete graph to another once every $\Theta(M)$ steps, and it must make $\Theta(M)$ such transitions, on average, to find the marked complete graph, resulting in $\Theta(M^2)$ total steps.

Next we analyze search on this by typical discrete-time and continuous-time quantum walks, highlighting that the number of walk steps per oracle query are different and must be accounted for in showing quadratic speedups over classical. Then we adjust the discrete-time quantum walk so that it can also take multiple walk steps per oracle query. This yields a cubic speedup over its corresponding classical walk with the same number of walk steps per oracle query, and is the first example of a greater-than-quadratic speedup.


\section{Discrete-Time Quantum Walks}

We begin with the typical discrete-time quantum walk search algorithm \cite{SKW2003}. The $N = M(M+1)$ vertices of the graph label computational basis states of an $N$-dimensional ``vertex'' Hilbert space, and the $M$ directions that a particle can move from each vertex spans an additional $M$-dimensional ``coin'' Hilbert space. Together, $\mathbb{C}^N \otimes \mathbb{C}^M$ is the Hilbert space of the system. Let $\ket{s_v}$ and $\ket{s_c}$ be uniform superpositions over the vertex and coin spaces, respectively:
\[ \ket{s_v} = \frac{1}{\sqrt{N}} \sum_{i=1}^N \ket{i}, \quad \ket{s_c} = \frac{1}{\sqrt{M}} \sum_{j=1}^M \ket{j}. \]
Then the system $\ket{\psi}$ begins in
\[ \ket{\psi_0} = \ket{s_v} \otimes \ket{s_c}, \]
which is an equal superposition over both the vertex and coin spaces. The discrete-time quantum walk is obtained by repeated applications of 
\[ U_0 = S \cdot ( I_N \otimes C_0 ), \]
where $C_0$ is the ``Grover diffusion'' coin \cite{SKW2003}
\[ C_0 = 2 \ketbra{s_c}{s_c} - I_M, \]
and $S$ is the ``flip-flop'' shift \cite{AKR2005} that causes the particle to hop and then turn around, \textit{e.g.}, $S (\ket{i} \otimes \ket{i \to j}) = \ket{j} \otimes \ket{j \to i}$.

With this choice of initial state $\ket{\psi_0}$ and evolution $U_0$, Fig.~\ref{fig:simplex_full} shows that there are only three types of vertices, which we indicate by identical colors and labels, each with two types of directions. In particular, the $a$ vertices can either point towards each other or towards the $b$ vertices, the $b$ vertices can either point towards the $a$ vertices or towards the $c$ vertices, and the $c$ vertices can either point towards the $b$ vertices or each other. So the system evolves in a 6D subspace, and we take uniform superpositions of identically evolving vertices/directions $\ket{aa}$, $\ket{ab}$, $\ket{ba}$, $\ket{bc}$, $\ket{cb}$, and $\ket{cc}$ as the basis vectors, \textit{e.g.},
\[ \ket{bc} = \frac{1}{\sqrt{M}} \sum_{b \in \text{blue}} \ket{b} \otimes \frac{1}{\sqrt{M-1}} \sum_{c \sim b} \ket{b \to c}. \]
In terms of these basis vectors, the initial state is
\begin{align*}
	\ket{\psi_0} = \frac{1}{\sqrt{N}} \Big( &\sqrt{M-1} \ket{aa} + \ket{ab} + \ket{ba} + \sqrt{M-1} \ket{bc} \\ &+ \sqrt{M-1} \ket{cb} + (M-1) \ket{cc} \Big),
\end{align*}
and the quantum walk operator is
\[ U_0 = \begin{pmatrix}
	\cos\theta & \sin\theta & 0 & 0 & 0 & 0 \\
	0 & 0 & -\cos\theta & \sin\theta & 0 & 0 \\
	\sin\theta & -\cos\theta & 0 & 0 & 0 & 0 \\
	0 & 0 & 0 & 0 & -\cos\theta & \sin\theta \\
	0 & 0 & \sin\theta & \cos\theta & 0 & 0 \\
	0 & 0 & 0 & 0 & \sin\theta & \cos\theta
\end{pmatrix}, \]
where $\cos\theta = 1-2/M$ and $\sin\theta = 2\sqrt{M-1}/M$.

Note that $U_0 \ket{\psi_0} = \ket{\psi_0}$, so to turn this quantum walk into a search algorithm, we use a different coin $C_1$ for the marked vertices and still use $C_0$ on the unmarked vertices. Then the search operator is
\[ U = S \cdot \left[ \left(I_N - \sum_w \ketbra{w}{w} \right) \otimes C_0 + \sum_w \ketbra{w}{w} \otimes C_1 \right]. \]
Since this distinguishes the marked from the unmarked vertices with each application of $U$, it performs one oracle query per walk step. Two choices for $C_1$ are common. The first is $C_1^{\text{flip}} = -C_0$, which causes $U$ to become
\begin{align*}
	U
	&= S \cdot \left[ \left(I_N - 2 \sum_w \ketbra{w}{w} \right) \otimes C_0 \right] \\
	&= \underbrace{S \cdot \left( I_N \otimes C_0 \right)}_{U_0} \cdot \underbrace{ \left(I_N - 2 \sum_w \ketbra{w}{w} \right) \otimes I_M }_{R_w}.
\end{align*}
Note $R_w = \text{diag}(-1,-1,1,1,1,1)$ is analogous to the phase flip in Grover's algorithm \cite{Grover1996,AKR2005,Wong2015b}. The second choice is $C_1^{\text{SKW}} = -I_M$, which gives search algorithms on the hypercube \cite{SKW2003} and arbitrary dimensional periodic grids \cite{AKR2005}. Even with these different coins for the marked vertices, the system still evolves in the same 6D subspace. 

\begin{figure}
\begin{center}
	\includegraphics{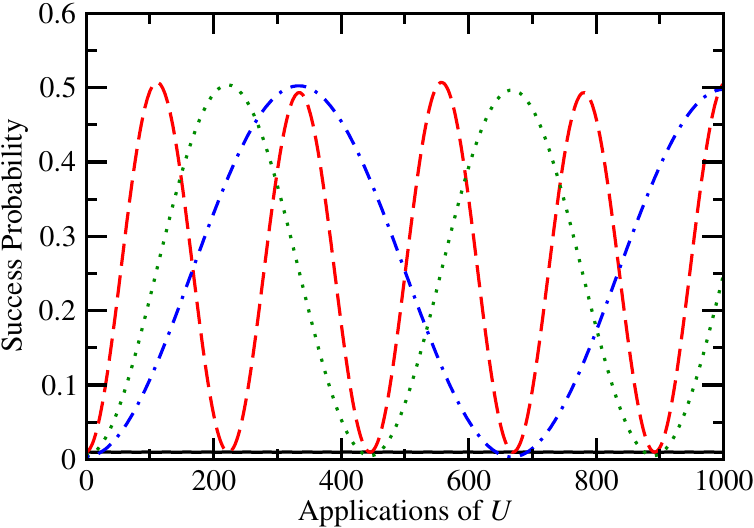}
	\caption{\label{fig:discrete_prob_time} The success probability as a function of the number of applications of $U$. The solid black line uses $C_1^{\text{flip}}$ with $M =100$, and the dashed red, dotted green, and dot-dashed blue curves use $C_1^{\text{SKW}}$ with $M = 100$, $200$, and $300$, respectively.}
\end{center}
\end{figure}

Figure~\ref{fig:discrete_prob_time} shows the success probability as we repeatedly apply $U$ with these two choices of $C_1$; with $C_1^{\text{flip}}$, the success probability stays near its initial value of $M/N = \Theta(1/M)$, and with $C_1^{\text{SKW}}$, the success probability reaches $1/2$ in $\Theta(M)$ steps. To prove these behaviors, we simply find the eigenvectors and eigenvalues of $U$ with each coin, the details of which are in Appendix~\ref{appendix:discrete}. With $C_1^\text{flip}$, the system approximately starts in an eigenstate. With $C_1^\text{SKW}$, it approximately starts as a linear combination of two eigenstates, and evolves to having equal probability of being in $\ket{ab}$ and $\ket{ba}$ (implying a success probability of $1/2$ from the $\ket{ab}$ term) after $\pi M / 2\sqrt{2}$ steps. While both of these are no better than the $\Theta(M)$ queries to guess which complete graph is marked, they are quadratically better than the $\Theta(M^2)$ queries needed by the classical walk that makes one query per walk step.


\section{Continuous-Time Quantum Walks}

\begin{figure}
\begin{center}
	\includegraphics{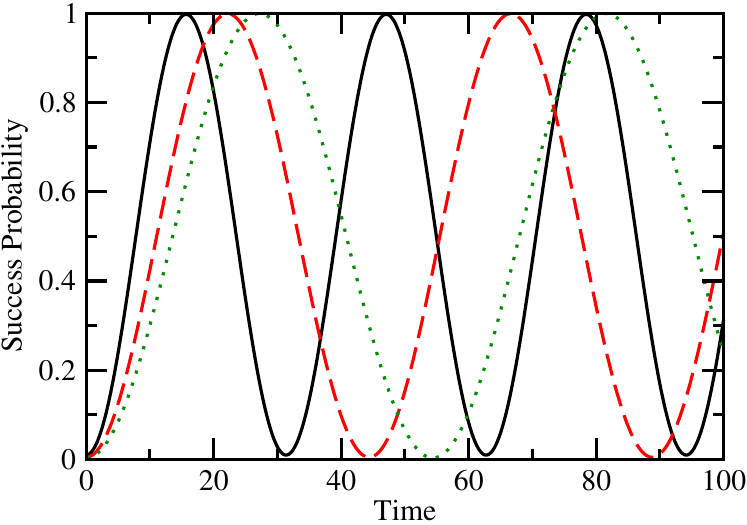}
	\caption{\label{fig:continuous_prob_time} The success probability as a function of time for continuous-time search with $\gamma = 1 + 1/M$. The solid black, dashed red, and dotted green curves are $M = 100$, $200$, and $300$, respectively.}
\end{center}
\end{figure}

Now let us consider search by continuous-time quantum walk, which does not require the coin space, so the system walks on the vertices labeling an $N$-dimensional Hilbert space. The system $\ket{\psi}$ begins in an equal superposition $\ket{s_v}$ over the vertices, and it evolves by Schr\"odinger's equation with Hamiltonian \cite{CG2004}
\begin{equation}
	\label{eq:H}
       	H = -\gamma A - \sum_w \ketbra{w}{w},
\end{equation}
where $\gamma$ is the jumping rate (\textit{i.e.}, amplitude per time), $A$ is the adjacency matrix of the graph ($A_{ij} = 1$ if $i$ and $j$ are adjacent and $0$ otherwise), and $w$ sums over the marked vertices. The first term effects a quantum walk \cite{CG2004} while the second term acts as an oracle \cite{Mochon2007}, with $\gamma$ setting their relative strength. As shown in Fig.~\ref{fig:simplex_full}, there are only three types of vertices. Since there are no directions, the system evolves in a 3D subspace spanned by uniform superpositions of the types of vertices:
\begin{gather*}
	\ket{a} = \frac{1}{\sqrt{M}} \sum_{i \in \text{red}} \ket{i}, \quad \ket{b} = \frac{1}{\sqrt{M}} \sum_{i \in \text{blue}} \ket{i} \\
	\ket{c} = \frac{1}{\sqrt{M(M-1)}} \sum_{i \in \text{white}} \ket{i}.
\end{gather*}
In this 3D basis, the search Hamiltonian \eqref{eq:H} is
\[ H = -\gamma \begin{pmatrix}
	M-1 + \frac{1}{\gamma} & 1 & 0 \\
	1 & 0 & \sqrt{M-1} \\
	0 & \sqrt{M-1} & M-1 \\
\end{pmatrix}. \]
Figure~\ref{fig:continuous_prob_time} shows the success probability as the system evolves by this Hamiltonian when $\gamma = 1 + 1/M$. As shown in Appendix~\ref{appendix:continuous} using a diagrammatic approach \cite{Wong2014} to degenerate perturbation theory \cite{JMW2014}, at this value of $\gamma$, two of the eigenstates of $H$ are approximately $(\ket{s_v} \pm \ket{a})/\sqrt{2}$ with eigenvalues $-M - 1 \mp 1/\sqrt{M}$, so the system evolves from $\ket{s_v}$ to $\ket{a}$ in time $\pi/\Delta E = \pi\sqrt{M}/2$.

While this appears to be a quartic speedup over the classical walk, it is not. The coefficient of the oracle term in the Hamiltonian \eqref{eq:H} is 1, so we are counting the number of oracle queries and allowing an arbitrary number of walk steps per oracle query. With $\gamma \approx 1$, this corresponds to $M$ walk steps per oracle query since the operator norm of $A$ is $M$ while the operator norm of the oracle term is $1$. Alternatively, if we use $\gamma A$ to define a classical random walk, it makes each transition with probability per time $\gamma$, and since there are $M$ possible transitions from each vertex, the probability of making some transition per time is $\gamma M$. A classical random walk that takes $M$ walk steps for every oracle query optimally finds a marked vertex with $\Theta(M)$ queries (but $\Theta(M^2)$ walk steps), and so the continuous-time algorithm is a quadratic speedup in the number of oracle queries.


\section{Multiple Walk Steps per Oracle Query}

Thus we have an example where the continuous-time quantum walk outperforms the discrete-time quantum walk in search, using the oracle for $\Theta(\sqrt{M})$ time rather than $\Theta(M)$ queries, in stark contrast to previous results showing that discrete-time quantum walks are faster \cite{AKR2005}. But it does not seem fair that the discrete-time quantum walk only take one walk step per oracle query when the continuous-time algorithm is allowed to take more. To provide greater equity, we modify the discrete-time algorithm to also take multiple walk steps per oracle query by repeatedly applying
\[ U = U_0^k R_w, \]
for some positive integer $k$, to the initial equal superposition over the coin and vertex spaces $\ket{\psi_0} = \ket{s_v} \otimes \ket{s_c}$. Then the number of applications of $U$ counts the number of oracle queries, similar to how evolution by \eqref{eq:H} counts the time that the oracle is used. A single application of $U$ acts on the initial state by
\begin{align*}
	U_0^k R_w \ket{\psi_0} 
	&= U_0^k \left[ \ket{\psi_0} - \frac{2}{\sqrt{N}} \left( \sqrt{M-1} \ket{aa} + \ket{ab} \right) \right] \\
	&= \ket{\psi_0} - \frac{2}{\sqrt{N}} U_0^k \left( \sqrt{M-1} \ket{aa} + \ket{ab} \right) \\
	&\approx \ket{\psi_0} - \frac{2\sqrt{M-1}}{\sqrt{N}} U_0^k \ket{aa},
\end{align*}
where we used $U_0 \ket{\psi_0} = \ket{\psi_0}$. If $U_0^k$ were not present, the second term would subtract amplitude from $\ket{aa}$, which would decrease the success probability. But we want to increase it instead, so we want to pick $k$ so as to add success amplitude. As shown in Appendix~\ref{appendix:multistep}, $\ket{aa} \approx ( \ket{\phi} + \ket{-\phi} )/\sqrt{2}$, where $\ket{\pm \phi}$ are eigenvectors of $U_0$ with eigenvalues $e^{\pm i\phi}$, where $\sin \phi \approx \sqrt{2/M}$. Then
\begin{align*}
	U_0^k R_w \ket{\psi_0} 
	&\approx \ket{\psi_0} - \frac{2\sqrt{M-1}}{\sqrt{N}} U_0^k \frac{1}{\sqrt{2}} \! \left( \ket{\phi} + \ket{-\phi} \right) \\
	&\approx \ket{\psi_0} - \frac{2\sqrt{M-1}}{\sqrt{N}} \frac{1}{\sqrt{2}} \! \left( e^{ik\phi} \ket{\phi} + e^{-ik\phi} \ket{-\phi} \right) \! .
\end{align*}
Then we pick
\[ k = \frac{(2n+1)\pi}{\phi} \approx \frac{(2n+1)\pi}{\sin\phi} \approx \frac{(2n+1)\pi\sqrt{M}}{\sqrt{2}}, \]
with integer $n$, so that the exponentials equal $-1$, flipping the second term so that it adds success amplitude rather than decreasing it:
\begin{align*}
	U_0^k R_w \ket{\psi_0} 
	&\approx \ket{\psi_0} + \frac{2\sqrt{M-1}}{\sqrt{N}} \frac{1}{\sqrt{2}} \! \left( \ket{\phi} + \ket{-\phi} \right) \\
	&\approx \ket{\psi_0} + \frac{2\sqrt{M-1}}{\sqrt{N}} \ket{aa}.
\end{align*}
With this choice of $k$, the algorithm takes $\Theta(\sqrt{M})$ walk steps for each oracle query.

\begin{figure}
\begin{center}
	\includegraphics{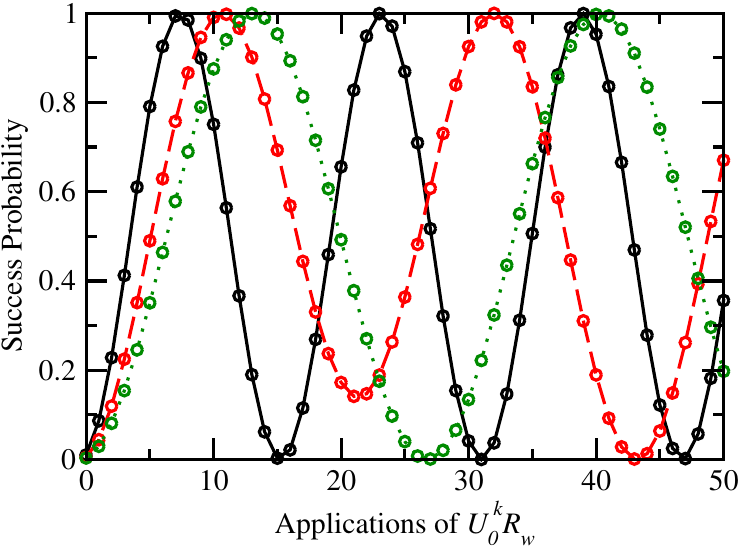}
	\caption{\label{fig:discrete_prob_time_multistep} The success probability as a function of the number of applications of $U_0^k R_w$ with $k = [ \pi\sqrt{M}/\sqrt{2} ]$. The solid black, dashed red, and dotted green curves are $M = 100$, $200$, and $300$, respectively.}
\end{center}
\end{figure}

As shown in Appendix~\ref{appendix:multistep}, with this choice of $k$, two of the eigenvectors of $U = U_0^k R_w$ are approximately $\ket{\pm \sigma} = (\ket{cc} \mp \ket{aa})/\sqrt{2}$ with eigenvalues $e^{\pm i \sigma}$, where $\sin \sigma \approx 2/\sqrt{M}$. Then the initial state $\ket{\psi_0}$ is approximately
\[ \ket{\psi_0} \approx \ket{cc} = \frac{1}{\sqrt{2}} \left( \ket{\sigma} + \ket{-\sigma} \right). \]
Acting on this $t$ times with $U_0^k R_w$,
\[ \left( U_0^k R_w \right)^t \ket{\psi_0} \approx \frac{1}{\sqrt{2}} \left( e^{i \sigma t} \ket{\sigma} + e^{-i \sigma t} \ket{-\sigma} \right). \]
When
\[ t = \frac{\pi}{2\sigma} \approx \frac{\pi}{2 \sin \sigma} \approx \frac{\pi \sqrt{M}}{4}, \]
this becomes
\[ \frac{1}{\sqrt{2}} \left( i \ket{\sigma} - i \ket{-\sigma} \right) = -i\ket{aa}. \]
So the system evolves from $\ket{\psi_0}$ to the marked vertices with probability $1$ in $\pi\sqrt{M}/4$ oracle queries, as shown in Fig.~\ref{fig:discrete_prob_time_multistep}.

A classical random walk that similarly takes $\Theta(\sqrt{M})$ walk steps for every oracle query finds a marked vertex with $\Theta(M^{3/2})$ queries (but $\Theta(M^2)$ walk steps). Thus our novel algorithm achieves a cubic speedup over the corresponding classical random walk with the same number of walk steps per oracle query, and it is the first example of a quantum walk search with a greater-than-quadratic speedup over its corresponding classical random walk. This contrasts with more general results with quadratic speedups \cite{Szegedy2004,KMOR2014}. Of course, if we allow the classical and quantum walks to take different numbers of walk steps per oracle query, then the speedup is quadratic, so specifying the number of walk steps per oracle query is critical in comparing algorithms.


\section{Conclusion}

We have shown that the typical discrete-time quantum walk search algorithm fixes the ratio of walk steps to oracle queries to 1-to-1, while the continuous-time algorithm allows for multiple walk steps per oracle query. When searching for a fully marked complete graph in the simplex of complete graphs, this results in the continuous-time algorithm outperforming the discrete-time's, evolving by the oracle for $\Theta(\sqrt{M})$ time instead of using $\Theta(M)$ queries. Both of these, however, are quadratic speedups over their corresponding classical algorithms. Providing greater equity, we modify the discrete-time algorithm to also allow multiple walk steps per oracle query. In doing so, the number of queries matches the $\Theta(\sqrt{M})$ runtime of the continuous-time algorithm, and it is a cubic speedup over the corresponding classical random walk.


\begin{acknowledgments}
	This work was supported by the European Union Seventh Framework Programme (FP7/2007-2013) under the QALGO (Grant Agreement No.~600700) project, and the ERC Advanced Grant MQC. 
\end{acknowledgments}


\appendix

\section{\label{appendix:discrete} Discrete-Time Algorithm}

\subsection{Phase Flip Coin}

With $C_1^\text{flip} = -C_0$, the evolution operator $U = U_0 R_w$ is
\[ U = \begin{pmatrix}
	-\cos\theta & -\sin\theta & 0 & 0 & 0 & 0 \\
	0 & 0 & -\cos\theta & \sin\theta & 0 & 0 \\
	-\sin\theta & \cos\theta & 0 & 0 & 0 & 0 \\
	0 & 0 & 0 & 0 & -\cos\theta & \sin\theta \\
	0 & 0 & \sin\theta & \cos\theta & 0 & 0 \\
	0 & 0 & 0 & 0 & \sin\theta & \cos\theta
\end{pmatrix}. \]
The eigenvalues of this are
\[ 1, -1, e^{i\phi_+}, e^{-i\phi_+}, e^{i\phi_-}, e^{-i\phi_-}, \]
where $\phi_\pm$ is defined such that
\begin{gather*}
	\cos\phi_\pm = \frac{\pm \sin\theta}{2} = \frac{\pm \sqrt{M-1}}{M}, \\
	\sin\phi_\pm = \frac{\sqrt{4-\sin^2\theta}}{2} = \frac{\sqrt{M^2-M+1}}{M}.
\end{gather*}
To find the eigenvectors, which have the form $\psi = (a,b,c,d,e,f)^\intercal$, we solve $U\psi = \lambda \psi$. This yields six equations:
\begin{gather*}
	-a \cos\theta - b \sin\theta = \lambda a \\
	-c \cos\theta + d \sin\theta = \lambda b \\
	-a \sin\theta + b \cos\theta = \lambda c \\
	-e \cos\theta + f \sin\theta = \lambda d \\
	c \sin\theta + d \cos\theta = \lambda e \\
	e \sin\theta + f \cos\theta = \lambda f.
\end{gather*}
Solving these yields
\begin{gather*}
	a = a, \quad b = \frac{-\lambda - \cos\theta}{\sin\theta} a, \quad c = \frac{-1 - \lambda \cos\theta}{\lambda \sin\theta} a, \\
	d = \frac{-\lambda^3 - \lambda^2 \cos\theta - \lambda \cos^2\theta - \cos\theta}{\lambda \sin^2\theta} a, \\
	e = \frac{-\lambda^3 \cos\theta - \lambda^2 \cos^2\theta - \lambda \cos\theta - 1}{\lambda^2 \sin^2\theta}a, \\
	f = \frac{-\lambda^3 \cos\theta - \lambda^2 \cos^2\theta - \lambda \cos\theta - 1}{\lambda^2 \sin\theta (\lambda-\cos\theta)} a.
\end{gather*}
Let $a = 1$. Then the (unnormalized) eigenvector corresponding to eigenvalue $\lambda = 1$ is
\begin{align*}
	\psi_1 = \bigg( &1, \frac{-1 - \cos\theta}{\sin\theta}, \frac{-1 - \cos\theta}{\sin\theta}, \frac{-(1+\cos\theta)^2}{\sin^2\theta}, \\
		 &\frac{-(1+\cos\theta)^2}{\sin^2\theta}, \frac{-(1+\cos\theta)^2}{\sin\theta(1-\cos\theta)} \bigg)^\intercal.
\end{align*}
Plugging in for $\cos\theta$ and $\sin\theta$, this becomes
\begin{align*}
	\psi_1 = \Big( &1, -\sqrt{M-1}, -\sqrt{M-1}, -(M-1), \\
			      &-(M-1), -(M-1)^{3/2} \Big)^\intercal.
\end{align*}
For large $M$, this is dominated by the last component, which means that the normalized eigenvector $\ket{\psi_1}$ is approximately $\ket{cc} \approx \ket{\psi_0}$. So the system approximately starts in an eigenstate and does not evolve significantly from it, as stated in the main text.


\subsection{SKW Coin}

With $C_1^\text{SKW} = -I_M$, the evolution operator $U$ is
\[ U = \begin{pmatrix}
		-1 & 0 & 0 & 0 & 0 & 0 \\
		0 & 0 & -\cos\theta & \sin\theta & 0 & 0 \\
		0 & -1 & 0 & 0 & 0 & 0 \\
		0 & 0 & 0 & 0 & -\cos\theta & \sin\theta \\
		0 & 0 & \sin\theta & \cos\theta & 0 & 0 \\
		0 & 0 & 0 & 0 & \sin\theta & \cos\theta
\end{pmatrix}. \]
The eigenvalues of this are
\[ -1, -1, e^{i\phi_+}, e^{-i\phi_+}, e^{i\phi_-}, e^{-i\phi_-}, \]
where $\phi_\pm$ is defined such that
\begin{align*}
	\cos\phi_\pm &= \frac{1 + \cos\theta \pm \alpha}{4} \\
			    &= \frac{M-1 \pm \sqrt{(M-1)(M+3)}}{2M}, \\
	\sin\phi_\pm &= \frac{\sqrt{2} \sqrt{5 - 2\cos\theta + \cos^2\theta \mp \alpha(1+\cos\theta)}}{4} \\
			    &= \frac{\sqrt{M^2 + 1 \mp (M-1)^{3/2}\sqrt{M+3}}}{\sqrt{2}M},
\end{align*}
where
\[ \alpha = \sqrt{(1+\cos\theta)(5-3\cos\theta)} = \frac{2}{M} \sqrt{(M-1)(M+3)}. \]
Clearly, one of the eigenvectors corresponding to eigenvalue $-1$ is $\ket{aa}$. The remaining eigenvectors have the form $\psi = (0,a,b,c,d,e)$. To find them, we solve $U \psi = \lambda \psi$, which yields five equations:
\begin{gather*}
	-b \cos\theta + c \sin\theta = \lambda a \\
	-a = \lambda b \\
	-d \cos\theta + e \sin\theta = \lambda c \\
	b \sin\theta + c \cos\theta = \lambda d \\
	d \sin\theta + e \cos\theta = \lambda e.
\end{gather*}
Solving these yields
\begin{gather*}
	a = -\lambda b, \quad b = b, \quad c = \frac{\cos\theta - \lambda^2}{\sin\theta} b, \\
	d = \frac{1 - \lambda^2 \cos\theta}{\lambda \sin\theta} b, \quad e = \frac{1 - \lambda^2 \cos\theta}{\lambda(\lambda - \cos\theta)} b.
\end{gather*}
Let $b = 1$. Then when $\lambda = e^{\pm i \phi}$, the (unnormalized) eigenvectors are
\begin{widetext}
	\[ \psi_{\pm \phi} = \begin{pmatrix}
			0 \\
			-\lambda \\
			1 \\
			\frac{\cos\theta - \lambda^2}{\sin\theta} \\
			\frac{1 - \cos\theta \lambda^2}{\lambda \sin\theta} \\
			\frac{1 - \cos\theta \lambda^2}{\lambda(\lambda - \cos\theta)}
		\end{pmatrix} = \begin{pmatrix}
			0 \\
			-\cos\phi \mp i \sin\phi \\
			1 \\
			\frac{\cos\theta - \cos2\phi}{\sin\theta} \mp i \frac{\sin2\phi}{\sin\theta} \\
			\frac{\cos\phi(1-\cos\theta)}{\sin\theta} \mp i \frac{\sin\phi(1+\cos\theta)}{\sin\theta} \\
			\frac{\cos2\phi - \cos\phi\cos\theta(1-\cos\theta) - \cos\theta}{1+\cos^2\theta-2\cos\phi\cos\theta} \mp i \frac{\sin2\phi-\sin\phi\cos\theta(1+\cos\theta)}{1+\cos^2\theta-2\cos\phi\cos\theta} 
	\end{pmatrix} . \]
\end{widetext}
Then the sum and difference (times $i$) of the eigenvectors with eigenvalues $e^{i\phi}$ and $e^{-i\phi}$ are
\begin{gather*}
	\psi_{+\phi} + \psi_{-\phi} = \begin{pmatrix}
		0 \\
		-2\cos\phi \\
		2 \\
		2\frac{\cos\theta - \cos2\phi}{\sin\theta} \\
		2\frac{\cos\phi(1-\cos\theta)}{\sin\theta} \\
		2\frac{\cos2\phi - \cos\phi\cos\theta(1-\cos\theta) - \cos\theta}{1+\cos^2\theta-2\cos\phi\cos\theta}
	\end{pmatrix}, \\
	i(\psi_{+\phi} - \psi_{-\phi}) = \begin{pmatrix}
		0 \\
		2 \sin\phi \\
		0 \\
		2 \frac{\sin2\phi}{\sin\theta} \\
		2 \frac{\sin\phi(1+\cos\theta)}{\sin\theta} \\
		2 \frac{\sin2\phi-\sin\phi\cos\theta(1+\cos\theta)}{1+\cos^2\theta-2\cos\phi\cos\theta} 
	\end{pmatrix}.
\end{gather*}
With $\phi_+$, these become for large $M$
\[ \psi_{+\phi_+} + \psi_{-\phi_+} \approx \begin{pmatrix} 
		0 \\
		-2 \\
		2 \\
		0 \\
		0 \\
		0 \\
	\end{pmatrix}, \quad
	i(\psi_{+\phi_+} - \psi_{-\phi_+}) \approx \begin{pmatrix} 
		0 \\
		0 \\
		0 \\
		0 \\
		0 \\
		2\sqrt{2} \\
\end{pmatrix}. \]
Normalizing, we see that the initial state $\ket{\psi_0}$ is approximately
\[ \ket{\psi_0} \approx \ket{cc} \approx \frac{i}{2\sqrt{2}} (\psi_{+\phi_+} - \psi_{-\phi_+}). \]
After $t$ applications of $U$, the system is approximately in the state
\[ U^t \ket{\psi_0} \approx \frac{i}{2\sqrt{2}} ( e^{i \phi_+ t} \psi_{+\phi_+} - e^{-i \phi_+ t}  \psi_{-\phi_+}) . \]
When $\phi_+ t = \pi/2$, \textit{i.e.}, when
\[ t = \frac{\pi}{2 \phi_+} \approx \frac{\pi}{2 \sin\phi_+} \approx \frac{\pi M}{2\sqrt{2}}, \]
then the state is
\begin{align*}
	\frac{i}{2\sqrt{2}} ( i \psi_{+\phi_+} - (-i) \psi_{-\phi_+})
	&= \frac{-1}{2\sqrt{2}} ( \psi_{+\phi_+} + \psi_{-\phi_+}) \\
	&= \frac{1}{\sqrt{2}} \begin{pmatrix} 
	0 \\
	1 \\
	-1 \\
	0 \\
	0 \\
	0 \\
\end{pmatrix}.
\end{align*}
So after $\pi M / 2\sqrt{2}$ applications of $U$, the system evolves from $\ket{s}$ to $\frac{1}{\sqrt{2}} (\ket{ab} - \ket{ba})$, which has probability $1/2$ of being at a marked vertex (from the $\ket{ab}$ piece), as stated in the main text.


\section{\label{appendix:continuous} Continuous-Time Algorithm}

\begin{figure}
	\begin{center}
		\subfloat[]{
			\includegraphics{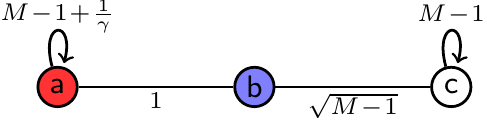}
			\label{fig:simplex_diagram_H}
		}

		\subfloat[]{
			\includegraphics{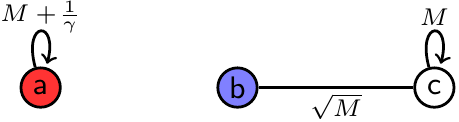}
			\label{fig:simplex_diagram_H0}
		}
		\caption{Apart from a factor of $-\gamma$, (a) the Hamiltonian for continuous-time search on the simplex of complete graphs with one fully marked complete graph, and (b) the leading-order terms.}
	\end{center}
\end{figure}

The search Hamiltonian is
\[ H = -\gamma \begin{pmatrix}
		M-1 + \frac{1}{\gamma} & 1 & 0 \\
		1 & 0 & \sqrt{M-1} \\
		0 & \sqrt{M-1} & M-1 \\
\end{pmatrix}. \]
To find the eigenvalues and eigenvectors of this, we employ degenerate perturbation theory \cite{JMW2014}. To begin, we visualize the Hamiltonian diagrammatically \cite{Wong2014} as a graph with three vertices, ignoring the overall factor of $-\gamma$, as shown in Fig.~\ref{fig:simplex_diagram_H}. We choose the leading-order Hamiltonian to exclude terms that scale less than $\sqrt{M}$, so it is:
\[ H^{(0)} = -\gamma \begin{pmatrix}
		M + \frac{1}{\gamma} & 0 & 0 \\
		0 & 0 & \sqrt{M} \\
		0 & \sqrt{M} & M \\
\end{pmatrix}. \]
Diagrammatically, we've eliminated edges that scale less than $\sqrt{M}$, as shown in Fig.~\ref{fig:simplex_diagram_H0}. From this diagram, we see that $H^{(0)}$ has three eigenvectors: one is $\ket{a}$, and the other two are linear combinations of $\ket{b}$ and $\ket{c}$. Specifically, they are
\begin{gather*}
	\ket{a}, \quad E = -\gamma \left( M + \frac{1}{\gamma} \right) \\
	\frac{-\sqrt{M} - \sqrt{M+4}}{2} \ket{b} + \ket{c}, \quad E = -\gamma \frac{M - \sqrt{M(M+4)}}{2} \\
	\frac{-\sqrt{M} + \sqrt{M+4}}{2} \ket{b} + \ket{c}, \quad E = -\gamma \frac{M + \sqrt{M(M+4)}}{2}
\end{gather*}
Note the third state is approximately
\begin{align*}
	\frac{1}{\sqrt{M}} \ket{b} + \ket{c} 
	&= \frac{1}{M} \sum_{i \in \text{blue}} \ket{i} + \frac{1}{\sqrt{M(M-1)}} \sum_{i \in \text{white}} \ket{i} \\
	       &\approx \frac{1}{M} \sum_{i \not\in \text{red}} \ket{i} = \ket{r},
\end{align*}
where $\ket{r}$ is the uniform superposition over the unmarked vertices. Setting the first and third eigenvalues equal (so that $\ket{a}$ and $\ket{r}$ are degenerate), we get the critical $\gamma$:
\[ \gamma_c = \frac{2}{-M + \sqrt{M(M+4)}} \approx \frac{M}{M-1} \approx 1 + \frac{1}{M}. \]

The perturbation $H^{(1)}$ restores terms of $\Theta(1)$, which in the diagram includes the edge of weight 1 between $\ket{a}$ and $\ket{b}$, the $-1$ in $\ket{a}$'s self-loop, and the $-1$ in $\ket{c}$'s self-loop. Away from the critical $\gamma$, the perturbation $H^{(1)}$ does not change the eigenvectors significantly, and so the initial state is approximately an eigenstate and fails to evolve beyond a global, unobservable phase. Near $\gamma_c$, however, the perturbation causes two linear combinations of $\ket{a}$ and $\ket{r}$,
\[ \ket{\psi} = \alpha_a \ket{a} + \alpha_r \ket{r}, \]
to be eigenstates of the perturbed system. The coefficients can be found by solving
\[ \begin{pmatrix}
		H_{aa} & H_{ar} \\
		H_{ra} & H_{rr} \\
	\end{pmatrix} \begin{pmatrix}
		\alpha_a \\
		\alpha_r \\
	\end{pmatrix} = E \begin{pmatrix}
		\alpha_a \\
		\alpha_r \\
\end{pmatrix}, \]
where $H_{ar} = \langle a | H^{(0)} + H^{(1)} | r \rangle$, \textit{etc}. With $\gamma = \gamma_c$, this yields
\[ \begin{pmatrix}
		-M-1 & \frac{-1}{\sqrt{M}} \\
		\frac{-1}{\sqrt{M}} & -M-1 \\ 
	\end{pmatrix} \begin{pmatrix}
		\alpha_a \\
		\alpha_r \\
	\end{pmatrix} = E \begin{pmatrix}
		\alpha_a \\
		\alpha_r \\
\end{pmatrix}, \]
for large $N$. Solving this, the eigenstates and eigenvalues of $H$ are approximately
\begin{gather*}
	\ket{\psi_0} = \frac{1}{\sqrt{2}} ( \ket{r} + \ket{a} ), \quad E = -M - 1 - \frac{1}{\sqrt{M}} \\
	\ket{\psi_1} = \frac{1}{\sqrt{2}} ( \ket{r} - \ket{a} ), \quad E = -M - 1 + \frac{1}{\sqrt{M}}.
\end{gather*}
Since $\ket{s_v} \approx \ket{r}$, these eigenstates are approximately $(\ket{s_v} \pm \ket{a})/\sqrt{2}$, as stated in the main text.


\section{\label{appendix:multistep} Multiple Walk Steps per Oracle Query}

Let us start by finding the eigenvalues and eigenstates of the quantum walk operator
\[ U_0 = \begin{pmatrix}
		\cos\theta & \sin\theta & 0 & 0 & 0 & 0 \\
		0 & 0 & -\cos\theta & \sin\theta & 0 & 0 \\
		\sin\theta & -\cos\theta & 0 & 0 & 0 & 0 \\
		0 & 0 & 0 & 0 & -\cos\theta & \sin\theta \\
		0 & 0 & \sin\theta & \cos\theta & 0 & 0 \\
		0 & 0 & 0 & 0 & \sin\theta & \cos\theta
\end{pmatrix}. \]
This has eigenvalues
\[ 1, -1, e^{i\phi_+}, e^{-i\phi_+}, e^{i\phi_-}, e^{-i\phi_-}, \]
where
\[ \cos\phi_\pm = \frac{\pm 1 + \cos\theta}{2}, \enspace \sin\phi_\pm = \frac{\sqrt{(1 \mp \cos\theta)(3 \pm \cos\theta)}}{2}. \]
To find the eigenstates, which have the form $\psi = (a,b,c,d,e,f)^\intercal$, we solve $U \psi = \lambda \psi$. This yields six equations:
\begin{gather*}
	a \cos\theta + b \sin\theta = \lambda a \\
	-c \cos\theta + d \sin\theta = \lambda b \\
	a \sin\theta - b \cos\theta = \lambda c \\
	-e \cos\theta + f \sin\theta = \lambda d \\
	c \sin\theta + d \cos\theta = \lambda e \\
	e \sin\theta + f \cos\theta = \lambda f .
\end{gather*}
Solving these yields
\begin{gather*}
	a = a, \quad b = \frac{\lambda - \cos\theta}{\sin\theta} a, \quad c = \frac{1 - \lambda \cos\theta}{\lambda \sin\theta} a, \\
	d = \frac{\lambda^3 - \lambda^2 \cos\theta - \lambda \cos^2\theta + \cos\theta}{\lambda \sin^2\theta} a, \\
	e = \frac{\lambda^3 \cos\theta - \lambda^2 \cos^2\theta - \lambda \cos\theta + 1}{\lambda^2 \sin^2\theta} a, \\
	f = \frac{\lambda^3 \cos\theta - \lambda^2 \cos^2\theta - \lambda \cos\theta + 1}{\lambda^2 \sin\theta (\lambda - \cos\theta)} a.
\end{gather*}
Let $a = 1$. Plugging in for $\lambda$, the (unnormalized) eigenvectors with eigenvalues $1$ and $-1$ are;
\begin{gather*}
	\psi_1 = \left( 1, \frac{1}{\sqrt{M-1}}, \frac{1}{\sqrt{M-1}}, 1, 1, \sqrt{M-1} \right)^\intercal \\
       	\psi_{-1} = \left( 1, -\sqrt{M-1}, -\sqrt{M-1}, 1, 1, \frac{-1}{\sqrt{M-1}} \right)^\intercal.
\end{gather*}
When $\lambda = e^{\pm i \phi}$, the (unnormalized) eigenvectors are
\begin{widetext}
	\[ \psi_{\pm \phi} = \begin{pmatrix}
			1 \\
			\frac{\cos\phi - \cos\theta}{\sin\theta} \pm i \frac{\sin\phi}{\sin\theta} \\
			\frac{\cos\phi - \cos\theta}{\sin\theta} \mp i \frac{\sin\phi}{\sin\theta} \\
			\frac{\cos2\phi - \cos^2\theta}{\sin^2\theta} \pm i \frac{\sin(2\phi) - 2\sin\phi\cos\theta}{\sin^2\theta} \\
			\frac{\cos2\phi - \cos^2\theta}{\sin^2\theta} \mp i \frac{\sin(2\phi) - 2\sin\phi\cos\theta}{\sin^2\theta} \\
			\frac{(1-2\cos2\phi)\cos\theta - \cos\phi\cos^2\theta + \cos3\phi + \cos^3\theta}{\sin\theta(1 - 2\cos\phi\cos\theta + \cos^2\theta)} \mp i \frac{\sin(3\phi) - 2\sin(2\phi)\cos\theta + \sin\phi\cos^2\theta}{\sin\theta(1 - 2\cos\phi\cos\theta + \cos^2\theta)} \\
	\end{pmatrix}. \]
\end{widetext}
Plugging in for $\phi_+$ and $\phi_-$, as well as for $\cos\theta$ and $\sin\theta$, this yields eigenvectors
\[ \psi_{\pm \phi_+} = \frac{1}{2} \begin{pmatrix}
		2 \\
		\frac{1}{\sqrt{M-1}} \pm i \frac{\sqrt{2M-1}}{\sqrt{M-1}} \\
		\frac{1}{\sqrt{M-1}} \mp i \frac{\sqrt{2M-1}}{\sqrt{M-1}} \\
		\frac{-1}{M-1} \pm i \frac{\sqrt{2M-1}}{M-1} \\
		\frac{-1}{M-1} \mp i \frac{\sqrt{2M-1}}{M-1} \\
		\frac{-2}{\sqrt{M-1}} \\
\end{pmatrix} \]
and
\[ \psi_{\pm \phi_-} = \frac{1}{2} \begin{pmatrix}
		2 \\
		-\sqrt{M-1} \pm i\sqrt{M+1} \\
		-\sqrt{M-1} \mp i\sqrt{M+1} \\
		-(M-1) \mp i\sqrt{M^2-1} \\
		-(M-1) \pm i\sqrt{M^2-1} \\
		2\sqrt{M-1} \\
\end{pmatrix}. \]
Let us take large $M$ so that $M+1 \approx M-1 \approx M$, $2M - 1 \approx 2M$, and $M^2 - 1 \approx M^2$. Then all the normalized eigenstates are approximately
\begin{align*}
	\ket{1} = \bigg( &\frac{1}{\sqrt{M}}, \frac{1}{M}, \frac{1}{M}, \frac{1}{\sqrt{M}}, \frac{1}{\sqrt{M}}, 1 \bigg)^\intercal \\
	\ket{-1} = \bigg( &\frac{1}{\sqrt{2M}}, \frac{-1}{\sqrt{2}}, \frac{-1}{\sqrt{2}}, \frac{1}{\sqrt{2M}}, \frac{1}{\sqrt{2M}}, \frac{-1}{\sqrt{2}M} \bigg)^\intercal \\
	\ket{\pm \phi_+} = \bigg( &\frac{1}{\sqrt{2}}, \frac{1}{2\sqrt{2M}} \pm \frac{i}{2}, \frac{1}{2\sqrt{2M}} \mp \frac{i}{2}, \\ &\frac{-1}{2\sqrt{2}M} \pm \frac{i}{2\sqrt{M}}, \frac{-1}{2\sqrt{2}M} \mp \frac{i}{2\sqrt{M}}, \frac{-1}{\sqrt{2M}} \bigg)^\intercal \\
	\ket{\pm \phi_-} = \bigg( &\frac{1}{M}, \frac{-1}{2\sqrt{M}} \pm \frac{i}{2\sqrt{M}}, \frac{-1}{2\sqrt{M}} \mp \frac{i}{2\sqrt{M}}, \\ &\frac{-1}{2} \mp \frac{i}{2}, \frac{-1}{2} \pm \frac{i}{2}, \frac{1}{\sqrt{M}} \bigg)^\intercal.
\end{align*}
Then
\[ \ket{aa} \approx \frac{1}{\sqrt{2}} \left( \ket{\phi_+} + \ket{-\phi_+} \right), \]
as stated in the main text.

Now let us find the eigenvalues and eigenvectors of $U^k R_w$ when $k = (2n+1)\pi\sqrt{M}/\sqrt{2}$ (which we assume from now on). Using the above eigenstates, we can define the matrices
\[ P = \begin{pmatrix}
		\ket{1} & \ket{-1} & \ket{\psi_+} & \ket{-\psi_+} & \ket{\psi_-} & \ket{-\psi_-}
\end{pmatrix} \]
and
\[ D^k = \text{diag} \left( 1, (-1)^k, -1, -1, e^{ik\phi_-}, e^{-ik\phi_-} \right) \]
that diagonalize $U_0$. Then
\[ U_0^k R_w = P D^k P^\dagger R_w \]
is, up to terms $O(1/\sqrt{M})$,
\begin{widetext}
	\[ \begin{pmatrix}
			1 & \frac{1+(-1)^k}{2 \sqrt{M}} & \frac{-1-(-1)^k}{2 \sqrt{M}} & 0 & 0 & \frac{2}{\sqrt{M}} \\
			\frac{1+(-1)^k}{2 \sqrt{M}} & \frac{1}{2} \left(1-(-1)^k\right) & \frac{1}{2} \left(1+(-1)^k\right) & \frac{-1-(-1)^k+2 \sin(k\phi_-)}{2 \sqrt{M}} & \frac{1-(-1)^k+2 \cos(k\phi_-)}{2 \sqrt{M}} & 0 \\
			\frac{1+(-1)^k}{2 \sqrt{M}} & \frac{1}{2} \left(-1-(-1)^k\right) & \frac{1}{2} \left(-1+(-1)^k\right) & \frac{1-(-1)^k+2 \cos(k\phi_-)}{2 \sqrt{M}} & \frac{-1-(-1)^k-2 \sin(k\phi_-)}{2 \sqrt{M}} & 0 \\
			0 & \frac{1+(-1)^k+2 \sin(k\phi_-)}{2 \sqrt{M}} & \frac{1-(-1)^k+2 \cos(k\phi_-)}{2 \sqrt{M}} & \cos(k\phi_-) & -\sin(k\phi_-) & \frac{1-\cos(k\phi_-)+\sin(k\phi_-)}{\sqrt{M}} \\
			0 & \frac{-1+(-1)^k-2 \cos(k\phi_-)}{2 \sqrt{M}} & \frac{-1 - (-1)^k + 2 \sin(k\phi_-)}{2 \sqrt{M}} & \sin(k\phi_-) & \cos(k\phi_-) & \frac{1-\cos(k\phi_-)-\sin(k\phi_-)}{\sqrt{M}} \\
			\frac{-2}{\sqrt{M}} & 0 & 0 & \frac{1-\cos(k\phi_-)-\sin(k\phi_-)}{\sqrt{M}} & \frac{1-\cos(k\phi_-)+\sin(k\phi_-)}{\sqrt{M}} & 1
	\end{pmatrix}. \]
\end{widetext}
To find the eigenvalues and eigenvectors of this, we use degenerate perturbation theory. The leading-order terms are
\begin{widetext}
	\[ (U_0^k R_w)^{(0)} \approx \begin{pmatrix}
			1 & 0 & 0 & 0 & 0 & 0 \\
			0 & \frac{1}{2} \left(1-(-1)^k\right) & \frac{1}{2} \left(1+(-1)^k\right) & 0 & 0 & 0 \\
			0 & \frac{1}{2} \left(-1-(-1)^k\right) & \frac{1}{2} \left(-1+(-1)^k\right) & 0 & 0 & 0 \\
			0 & 0 & 0 & \cos(k\phi_-) & -\sin(k\phi_-) & 0 \\
			0 & 0 & 0 & \sin(k\phi_-) & \cos(k\phi_-) & 0 \\
			0 & 0 & 0 & 0 & 0 & 1
\end{pmatrix}. \]
\end{widetext}
Let us separately consider the cases when $k$ is even and when $k$ is odd.

When $k$ is even, the leading-order terms of $U_0^k R_w$ are
\[ (U_0^k R_w)^{(0)}_{k \text{ even}} \approx \begin{pmatrix}
	1 & 0 & 0 & 0 & 0 & 0 \\
	0 & 0 & 1 & 0 & 0 & 0 \\
	0 & -1 & 0 & 0 & 0 & 0 \\
	0 & 0 & 0 & \cos(k\phi_-) & -\sin(k\phi_-) & 0 \\
	0 & 0 & 0 & \sin(k\phi_-) & \cos(k\phi_-) & 0 \\
	0 & 0 & 0 & 0 & 0 & 1
\end{pmatrix}, \]
which has eigenvectors and eigenvalues
\begin{gather*}
	\frac{1}{\sqrt{2}} \left( -i \ket{ab} + \ket{ba} \right), \quad i \\
	\frac{1}{\sqrt{2}} \left( i \ket{ab} + \ket{ba} \right), \quad -i \\
	\ket{aa}, \quad 1 \\
	\ket{cc}, \quad 1 \\
	\frac{1}{\sqrt{2}} \left( -i \ket{bc} + \ket{cb} \right), \quad e^{-ik\phi_-} \\
	\frac{1}{\sqrt{2}} \left( i \ket{bc} + \ket{cb} \right), \quad e^{ik\phi_-} .
\end{gather*}
The eigenstates we care about are $\ket{aa}$ and $\ket{cc}$, which are degenerate. The perturbation $(U_0^k R_w)^{(1)}$ restores terms of $\Theta(1/\sqrt{M})$ in the matrix, and two linear combinations of $\ket{aa}$ and $\ket{cc}$,
\[ \alpha_a \ket{aa} + \alpha_c \ket{cc}, \]
are eigenstates of the perturbed system. The coefficients can be found by solving
\[ \begin{pmatrix}
	H_{aa} & H_{ac} \\
	H_{ca} & H_{cc} \\
\end{pmatrix} \begin{pmatrix}
	\alpha_a \\
	\alpha_c \\
\end{pmatrix} = E \begin{pmatrix}
	\alpha_a \\
	\alpha_c \\
\end{pmatrix}, \]
where $H_{ac} = \langle aa | (U_0^k R_w)^{(0)} + (U_0^k R_w)^{(1)} | cc \rangle$, etc. This yields
\[ \begin{pmatrix}
	1 & \frac{2}{\sqrt{M}} \\
	\frac{-2}{\sqrt{M}} & 1 \\
\end{pmatrix} \begin{pmatrix}
	\alpha_a \\
	\alpha_c \\
\end{pmatrix} = E \begin{pmatrix}
	\alpha_a \\
	\alpha_c \\
\end{pmatrix}. \]
Solving this, the eigenvectors and eigenvalues of $U_0^k R_w$ for even $k$ are
\[ \ket{\pm \sigma} = \frac{1}{\sqrt{2}} \left( \mp \ket{aa} + \ket{cc} \right), \quad E_\pm = 1 \pm \frac{2}{\sqrt{M}} i \approx e^{\pm i \sigma}, \]
where $\sin\sigma \approx 2/\sqrt{M}$, as stated in the main text.

What about when $k$ is odd? We get
\[ (U_0^k R_w)^{(0)}_{k \text{ odd}} \approx \begin{pmatrix}
	1 & 0 & 0 & 0 & 0 & 0 \\
	0 & 1 & 0 & 0 & 0 & 0 \\
	0 & 0 & -1 & 0 & 0 & 0 \\
	0 & 0 & 0 & \cos(k\phi_-) & -\sin(k\phi_-) & 0 \\
	0 & 0 & 0 & \sin(k\phi_-) & \cos(k\phi_-) & 0 \\
	0 & 0 & 0 & 0 & 0 & 1
\end{pmatrix}, \]
which has eigenvectors and eigenvalues
\begin{gather*}
	\ket{ba}, \quad -1 \\
	\ket{aa}, \quad 1 \\
	\ket{ab}, \quad 1 \\
	\ket{cc}, \quad 1 \\
	\frac{1}{\sqrt{2}} \left( -i\ket{bc} + \ket{cb} \right), \quad e^{-ik\phi_-} \\
	\frac{1}{\sqrt{2}} \left( i\ket{bc} + \ket{cb} \right), \quad e^{ik\phi_-}.
\end{gather*}
The three eigenstates we care about are the degenerate states $\ket{aa}$, $\ket{ab}$, and $\ket{cc}$. With the $\Theta(1/\sqrt{M})$ perturbation $(U_0^k R_w)^{(1)}$, three superpositions of these states $\alpha_a \ket{aa} + \alpha_b \ket{ab} + \alpha_c \ket{cc}$ become eigenstates of the perturbed system, where the coefficients can be found by solving
\[ \begin{pmatrix}
	H_{aa} & H_{ab} & H_{ac} \\
	H_{ba} & H_{bb} & H_{bc} \\
	H_{ca} & H_{cb} & H_{cc} \\
\end{pmatrix} \begin{pmatrix}
	\alpha_a \\
	\alpha_b \\
	\alpha_c \\
\end{pmatrix} = E \begin{pmatrix}
	\alpha_a \\
	\alpha_b \\
	\alpha_c \\
\end{pmatrix}, \]
where $H_{ab} = \langle aa | (U_0^k R_w)^{(0)} + (U_0^k R_w)^{(1)} | ab \rangle$, \textit{etc}. This yields
\[ \begin{pmatrix}
	1 & 0 & \frac{2}{\sqrt{M}} \\
	0 & 1 & 0 \\
	\frac{-2}{\sqrt{M}} & 0 & 1 \\
\end{pmatrix} \begin{pmatrix}
	\alpha_a \\
	\alpha_b \\
	\alpha_c \\
\end{pmatrix} = E \begin{pmatrix}
	\alpha_a \\
	\alpha_b \\
	\alpha_c \\
\end{pmatrix}. \]
Solving this, the eigenvectors are
\[ \ket{ab}, \quad 1 \]
and
\[ \ket{\pm \sigma} = \frac{1}{\sqrt{2}} \left( \mp \ket{aa} + \ket{cc} \right), \quad 1 \pm \frac{2}{\sqrt{M}} i \approx e^{\pm i \sigma} \]
where $\sin\sigma \approx 2/\sqrt{M}$. So we get the same result as for even $k$. While not relevant for our results, note that the term $\ket{ab}$ can have an affect after the first peak in success probability, causing a double peak, which the analytics reveal when higher-order terms are included in the calculation.


\bibliography{refs}

\end{document}